\documentclass[prb,amsmath,amssymb,twocolumn,showpacs,floatfix]{revtex4}
\usepackage{graphicx}
\usepackage{dcolumn}
\usepackage{bm}
\newcommand{\ve}[1]{\mbox{\boldmath$#1$}}

\begin{document}

\preprint{APS/123-QED}
\title{Muon spin rotation measurements of the vortex state in vanadium: A comparative analysis using iterative 
and analytical solutions of the Ginzburg-Landau equations}

\author{M.~Laulajainen,$^{1}$ F.D.~Callaghan,$^{1}$ C.V.~Kaiser,$^{1}$ and J.E.~Sonier$^{1,2,}$}
\email{jsonier@sfu.ca}
\affiliation{$^1$Department of Physics, Simon Fraser University, Burnaby, British Columbia V5A 1S6, Canada \\
$^2$Canadian Institute for Advanced Research, 180 Dundas Street West, Toronto, Ontario M5G 1Z8, Canada} 

\date{\today}

\begin{abstract}
We report muon spin rotation measurements on a single crystal of the marginal 
type-II superconductor V. The measured internal magnetic field distributions
are modeled assuming solutions of the Ginzburg-Landau (GL) equations for an
ideal vortex lattice obtained using (i) an iterative procedure developed by 
E.H.~Brandt, Phys.~Rev.~Lett. {\bf 78}, 2208 (1997) and (ii) 
a variational method. 
Both models yield qualitatively similar results.
The magnetic penetration depth $\lambda$ and the GL coherence length $\xi$ 
determined from the data 
analysis exhibit strong field dependences, which are attributed to changes in 
the electronic structure of the vortex lattice. The zero-field extrapolated
values of $\lambda$ and the GL parameter $\kappa$ agree well with
values obtained by other experimental techniques that probe the Meissner state.       
\end{abstract}

\maketitle

\section{Introduction}
In order to analyze muon spin rotation ($\mu$SR) measurements on a type-II 
superconductor in the vortex state, it is necessary to assume a theoretical model 
for the spatial variation of the local internal magnetic field $B({\bf r})$.\cite{SonierRMP}  
An essential requirement of the model is that it must account for the finite size of 
the vortex cores. Thus far the internal magnetic field distribution $n(B)$ measured
by $\mu$SR has been analyzed assuming analytical models for $B({\bf r})$
based on London and Ginzburg-Landau (GL) theories.
Since London theory does not account for the finite size of the vortex cores,
a cutoff factor derived from GL theory must be inserted into the analytical
London expression for       
$B({\bf r})$ to correct for the divergence of $B({\bf r})$ as $r \! \rightarrow \! 0$.
Unfortunately, analytical cutoff factors are derivable from GL theory only near the lower and upper
critical fields, $B_{c1}$ and $B_{c2}$. At intermediate fields, these analytical cutoffs deviate
substantially from the precise numerical calculations,\cite{Oliveira:98} making
modified London models inappropriate for the analysis of $\mu$SR data. 
There are several approximate analytical expressions for $B({\bf r})$ that have been derived 
from the GL equations.\cite{Clem:75,Yaouanc:97, Hao:91, Pogosov:01} 
For example, a variational solution 
of the GL equations \cite{Clem:75,Yaouanc:97} has proven to be a reliable model for analyzing
$\mu$SR  measurements on V$_3$Si (Ref.\cite{Sonier:04}), NbSe$_2$ (Ref.\cite{Callaghan:05}) and 
YBa$_2$Cu$_3$O$_{7-\delta}$ (Ref.\cite{Sonier:97,Sonier:99}).  However, this analytical
GL model is strictly valid only at low reduced fields $b \! = \! B/B_{c2}$ and
large values of the GL parameter $\kappa$. Thus,   
often used analytical models for $B({\bf r})$ have limited validity and can 
deviate substantially from the numerical solutions of the GL equations.

Brandt has developed an iterative method for solving the GL equations that accurately determines
$B({\bf r})$ for arbitrary $b$, $\kappa$, and vortex-lattice symmetry.\cite{Brandt:97} Thus far
this iteration method has not been applied to the analysis of $\mu$SR measurements of $n(B)$ 
in the vortex state. As a first test of this method we have chosen to study the marginal 
type-II superconductor vanadium (V). This rigorous analysis method is expected to be required for V, 
whose low value of $\kappa$ falls outside the range of validity of 
the analytical models. 
In addition, the low value of $B_{c2}$ ($\approx 0.45$~T) gives us experimental access to reduced 
fields which are beyond the range of validity of the analytical model.

The paper is organized as follows: Theoretical models for $B({\bf r})$ are described in Sec.
II. The experimental procedures are described in Sec. III. Measurements in zero external
magnetic field are presented in Sec. IV. Measurements in the vortex 
state are described in Sec. V and concluding remarks are given in Sec. VI.  
 
\section{Theoretical models}
\subsection{Iterative GL solution}
\begin{figure}
\centering
\includegraphics[width=9.0cm]{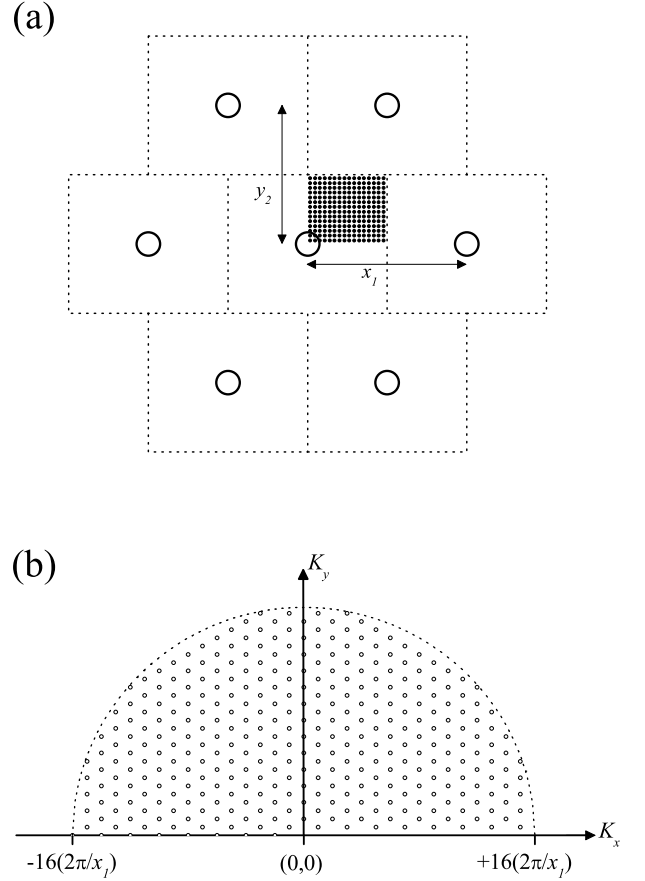}
\caption{(a) The vortex positions in the hexagonal lattice ($\bigcirc$) and the rectangular unit 
cell (dotted lines). Due to symmetry, the theoretical probability field distribution $n(B)$
was calculated at approximately 950 locations ($\bullet$) within just one-quarter of the 
rectangular unit cell. For the sake of clarity, the sampled positions are shown at 1/4 of the 
density used in the calculations.
(b) The points in the reciprocal space lattice ($\circ$) that are included in the ${\bf K}$ sums
of Eqs.~(\ref{omega})--(\ref{Q}), Eq.~(\ref{modLondon}) and Eq.~(\ref{analyticalGL}).}
\label{cells}
\end{figure}

Here we briefly outline the iteration method presented in Ref.~\cite{Brandt:97}, and correct 
some typographical errors contained therein.
The GL equations are written in terms of the real order parameter $\omega$, the local 
magnetic field $B$, and the supervelocity ${\bf Q}$, which are expressed as the Fourier series
\begin{equation}
\omega (\ve{\mathrm{r}})=\sum_{\bf K}a_{\bf K}(1-\cos\ve{\mathrm{K}}\cdot\ve{\mathrm{r}}),
\label{omega}
\end{equation}

\begin{equation}
B(\ve{\mathrm{r}})=\bar{B} + \sum_{\bf K}b_{\bf K}\cos\ve{\mathrm{K}}\cdot\ve{\mathrm{r}},
\label{B}
\end{equation}

\begin{equation}
\ve{\mathrm{Q}}(\ve{\mathrm{r}})=\ve{\mathrm{Q}}_A(\ve{\mathrm{r}}) + \sum_{\bf K}b_{\bf K}
\frac{\hat{\ve{\mathrm{z}}}\times \ve{\mathrm{K}}}{K^2}\sin\ve{\mathrm{K}}\cdot\ve{\mathrm{r}},
\label{Q}
\end{equation}
where $a_{\bf K}$ and $b_{\bf K}$ are Fourier coefficients, $\omega({\bf r})=|\psi({\bf r})|^2$, 
$\bar{B}$ is the average internal field, 
and $\psi({\bf r})$ is the complex GL order parameter. The ``tail'' of the position vector 
${\bf r} \! = \! (x, y)$ is at the vortex center $(0, 0)$. 
The local magnetic field $B$ is given in units of
$\sqrt{2} B_c$ (where $B_c$ is the thermodynamic critical field)
and all length scales are in units of the magnetic penetration depth $\lambda$.
$\ve{\mathrm{Q}}_A(\ve{\mathrm{r}})$ is the supervelocity obtained from
Abrikosov's solution of the GL equations near $B_{c2}$
\begin{equation}
\ve{\mathrm{Q}}_A({\ve{\mathrm{r}}})=\frac{{\bf {\nabla}} \omega_A \times \hat{\ve{\mathrm{z}}}}{2\kappa
\omega_A},
\end{equation}
where $\omega_A({\bf r})$ is calculated from Eq.~(\ref{omega}) using
\begin{equation}
a_{\bf K}^A=-(-1)^\nu \exp(-\pi \nu \sqrt{3}).
\end{equation}
Here $\nu=m^2+mn+n^2$, assuming a hexagonal vortex lattice with vortex positions given by 
\begin{equation}
\ve{\mathrm{R}}_{\mathrm{mn}}=(mx_1+nx_2,ny_2),
\end{equation}
where $m$ and $n$ are integers, $x_1$ is the intervortex spacing, $x_2=x_1/2$ and $y_2=x_1\sqrt{3}/2$.
The spatial field profile $B({\bf r})$ was calculated at approximately 950 locations within one-quarter 
of the rectangular unit cell shown in Fig.~\ref{cells}(a).
The reciprocal lattice vectors used in the calcluation of $B({\bf r})$ are given by 
\begin{equation}
\ve{\mathrm{K}}\equiv\ve{\mathrm{K}}_{\mathrm{mn}}=\frac{2\pi}{S}(my_2,nx_1+mx_2),
\end{equation}
where $S$ is the unit cell area.
The $\ve{\mathrm{K}}_{\mathrm{mn}}$ vectors are restricted to those indicated in 
Fig.~\ref{cells}(b), corresponding to $-16\!\ge\!m\!\le\!16$ and $-16\!\ge\!n\!\le\!16$ 
within a semicircle with $K_y \ge 0$ 
(but excluding vectors with $K_x\!\ge\!0$ and $K_y\!=\!0$).  
It was found that the calculated field distribution did not change 
significantly if the summation was extended to values of $|n|$ and $|m|$ greater than 16. 

The Fourier coefficients $a_{\bf K}$ and $b_{\bf K}$ are 
calculated from 
\begin{equation}
a_{\bf K}=\frac{4\kappa^2<(\omega^2-2\omega+\omega Q^2+g)
\cos\ve{\mathrm{K}}\cdot\ve{\mathrm{r}}>}{K^2+2\kappa^2},
\label{ak1}
\end{equation}

\begin{equation}
a_{\bf K}=a_{\bf K}\cdot<\omega-\omega Q^2-g>/<\omega^2>,
\label{ak2}
\end{equation}

\begin{equation}
b_{\bf K}=\frac{-2<[\omega B+\bar{\omega}(\bar{B}-B)+p]
\cos\ve{\mathrm{K}}\cdot\ve{\mathrm{r}}>}{K^2+\bar{\omega}},
\label{bk}
\end{equation}
where 
$g\!=\!(\nabla\omega)^2/(4\kappa^2\omega)$,
$\bar{\omega}$ is the spatial average of $\omega$
and
$p\!=\!(\nabla \omega \times \ve{\mathrm{Q}})\hat{\ve{\mathrm{z}}}\!=\!Q_y\frac{\partial \omega}
{\partial x}-Q_x\frac{\partial \omega}{\partial y}.$
We note that the definitions of $p$ and the $\bar{\omega}(\bar{B}-B)$ term in 
Eq.~(\ref{bk}) are incorrectly written in the original article,\cite{Brandt:05} but have 
been corrected here.
As explained in Ref.~\cite{Brandt:97}, solutions to the GL equations are acquired by 
first iterating Eqs.~(\ref{omega}), (\ref{ak1}) and (\ref{ak2}) a few 
times to relax $\omega$ and then iterating Eqs. (\ref{bk}), (\ref{omega}),
(\ref{B}), (\ref{Q}), (\ref{ak1}), (\ref{ak2}) and again (\ref{bk})
etc\dots~to relax $B$.

\subsection{Comparison with other models}
Here we compare the results of the above iteration method for $B({\bf r})$
to the widely used modified London and analytical GL models. The local magnetic
field at position ${\bf r} \! = \! (x, y)$ in the modified London 
model\cite{Brandt:88} is 
\begin{equation}
B(\ve{\mathrm{r}})=\bar{B}\sum_{\bf K}
\frac{e^{-i\ve{\mathrm{K}}\cdot \ve{\mathrm{r}}}
e^{-K^2\xi^2/2}}{1+\lambda^2K^2}.
\label{modLondon}
\end{equation}
Although this model is considered applicable for reduced fields 
$b \! = \! B/B_{c2} \! \leq \! 0.25$ and 
$\kappa \! \geq \! 2$, the Gaussian cutoff factor $\exp(-K^2\xi^2/2)$
introduced to account for the logarithmic divergence of $B({\bf r})$ at
the center of the vortex is not strictly valid.\cite{Oliveira:98}

The approximate analytical solution of the GL equations for $B({\bf r})$
is\cite{Yaouanc:97} 
\begin{equation}
B(\ve{\mathrm{r}})=\bar{B}(1-b^4)\sum_{\bf K}
\frac{e^{-i\ve{\mathrm{K}}\cdot \ve{\mathrm{r}}}uK_1(u)}{\lambda^2K^2},
\label{analyticalGL}
\end{equation}
where
\begin{equation}
u^2=2\xi^2K^2(1+b^4)[1-2b(1-b)^2].
\end{equation}
$K_1$ is a modified Bessel function and $\xi$ is the
GL coherence length. This analytical GL model is a reasonable
approximation for $b \! \ll \! 1$ and $\kappa \! \gg \! 1$.

\begin{figure}
\centering
\includegraphics[width=6.0cm, angle=270]{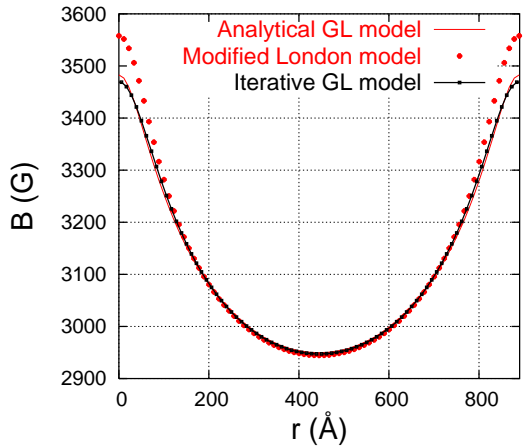}
\caption{(Color online) The spatial field profile $B(r)$ along the straight line connecting 
nearest-neighbor vortices, for an hexagonal vortex lattice, $\kappa\!=\!25$, 
$\bar{B} \! = \!3$~kG and $\lambda\!=\!$1000~{\AA}.}
\label{v3si}
\end{figure}

\begin{figure}
\centering
\includegraphics[width=6.0cm, angle=270]{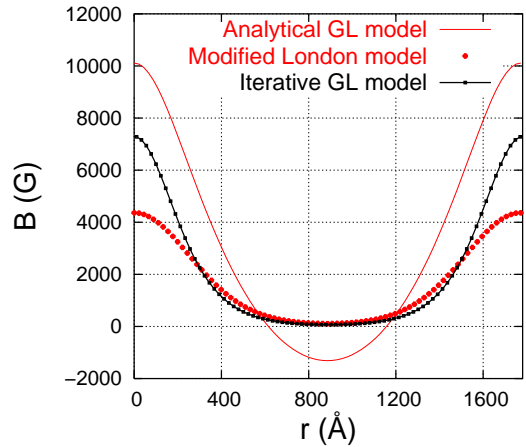}
\caption{(Color online) The spatial field profile $B(r)$ along the straight line connecting 
nearest-neighbor vortices, for an hexagonal vortex lattice, $\kappa\!=\!1/\sqrt{2}$, 
$\bar{B}\!=\!753$~G and $\lambda\!=\!$150~{\AA}.}
\label{sqrt}
\end{figure}

\begin{figure}
\centering
\includegraphics[width=6.0cm, angle=270]{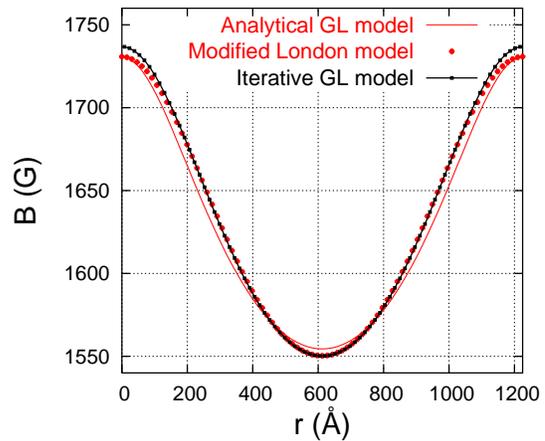}
\caption{(Color online) The spatial field profile $B(r)$ along the straight line connecting 
nearest-neighbor vortices, for an hexagonal vortex lattice, 
$\kappa\!=\!5.3$, $\bar{B}\!=\!1.6$~kG 
and $\lambda\!=\!$1042~{\AA}.}
\label{vanad}
\end{figure}

Figures \ref{v3si}, \ref{sqrt} and \ref{vanad} show comparisons between the
solutions for $B({\bf r})$ from the three different models, plotted along the straight
line connecting nearest-neighbor vortices. The parameters used to generate the 
curves in Fig.~\ref{v3si} are characteristic of the high-$\kappa$
superconductor V$_3$Si (Ref.\cite{Sonier:04}). While there is good agreement between
the iterative and analytical solutions of the GL equations, the modified London model
deviates substantially in the region of the vortex cores.
Figure~\ref{sqrt} shows that the modified London and analytical GL models completely 
break down for $\kappa \! = \! 1/\sqrt{2}$, the limit of type-II superconductivity.  
For example, in the case of the analytical GL model, there is actually
a region between the vortices where the local field changes direction.
In Fig.~\ref{vanad}, plots of $B({\bf r})$ are shown for a set of parameters 
obtained from $\mu$SR measurements on the low-$\kappa$ superconductor V 
(see Sec.~V). The value $\kappa \! = \! 5.3$ is rather large for V,
but as we explain in Sec.~V, $\kappa$ is really an ``effective'' fit parameter
influenced by the electronic structure of the vortex lattice.  

In Fig.~\ref{vanad} the analytical GL solution deviates from the iterative GL 
solution both near and far from the vortex core regions, whereas the modified London
model fails only in the region of the vortex cores.

\section{Experimental details}
The single crystal of the low-$\kappa$ superconductor V measured in 
the present study was purchased from Goodfellow Cambridge Ltd.\cite{Goodfellow}
The crystal is a disk, 13~mm in diameter by 0.35~mm thick, with the 
$\left<111\right>$ crystallographic direction perpendicular to the plane of the disk.
Magnetization measurements indicate that the crystal has a superconducting
transition temperature of $T_c \! = \! 5.2$~K, and an upper critical field
of $H_{c2} \approx 4.2$~kOe. This value of $H_{c2}$ corresponds to a BCS coherence length 
of $\xi_0 \approx 280$~{\AA}. 
A four-probe potentiometric ac resistivity measurement yielded 
$\rho = 0.7$~$\mu\Omega\cdot$cm just above $T_c$, which is 31 times smaller than $\rho$ 
at rrom temperature.
Using the carrier concentration $n \approx 9 \times 10^{28}$~m$^{-3}$ obtained 
from Hall resistance measurements,\cite{Hurd} a free electron theory\cite{Kittel}
calculation of the mean free path yields $l = \hbar k_F/\rho n e^2 \approx 900$~{\AA},  
where $k_F = (3 \pi^2 n)^{1/3}$ is the Fermi wavenumber and $e$ is the electronic charge.
Thus our sample is in the clean limit with $l / \xi_0 \approx 3$.
Neutron scattering measurements performed on our V 
single crystal showed no evidence of an intermediate mixed state ({\it i.e.} a mixture 
of Meissner and vortex-lattice phases).\cite{Forgan:05}

The $\mu$SR measurements were carried out on the M15 beamline at the 
Tri-University Meson Facility (TRIUMF), Vancouver, Canada, using a
dilution refrigerator to cool the sample.   
Measurements of the vortex state were done under field-cooled conditions
in a ``transverse field'' (TF) geometry, in 
which the magnetic field was applied along the $z$-axis parallel to the $\left<111\right>$ 
direction of the crystal, but perpendicular to the initial muon spin polarization $P_x(0)$
(which defines the $x$-axis). Each measurement was done by implanting approximately 
$2 \! \times \! 10^7$ spin-polarized  muons one at a time into the crystal, where their spins
precess around the local magnetic field $B({\bf r})$ at the Larmor 
frequency $\omega = \gamma_{\mu} B$, where $\gamma_\mu \! = \! 0.0852$~$\mu$s$^{-1}$~G$^{-1}$
is the muon gyromagnetic ratio. The muons stop randomly on the length scale 
of the vortex lattice, and hence evenly sample $B({\bf r})$. 
The $\mu$SR signal obtained by the detection of the decay positrons from 
an ensemble of muons implanted into the single crystal is given by
\begin{equation}
A(t)  = a_0 P_x(t) \,  ,
\label{eq:asy} 
\end{equation}
where $A(t)$ is called the $\mu$SR ``asymmetry'' spectrum, $a_0$ is the asymmetry maximum, 
and $P_x(t)$ is the time evolution of the muon spin polarization
\begin{equation}
P_x(t) = \int_0^{\infty} n(B) \cos(\gamma_\mu B t + \phi) dB \, .
\label{eq:polarization}  
\end{equation}
Here $\phi$ is a phase constant, and
\begin{equation}
n(B^{\prime}) = \langle \delta [ B^{\prime} - B({\bf r})] \rangle \, ,
\end{equation}
is the probability of finding a local field $B$ in the $z$-direction at an arbitrary position {\bf r} 
in the $x$-$y$ plane. Further details on this application of the $\mu$SR technique 
are found in Ref.~\cite{SonierRMP}.

\begin{figure}
\centering
\includegraphics[width=9.0cm]{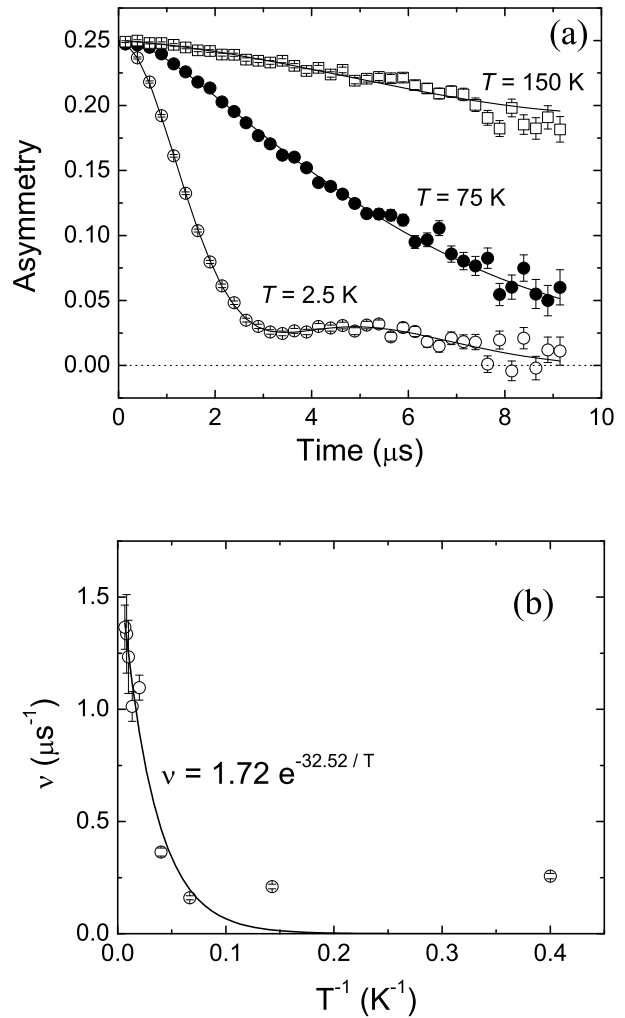}
\caption{(a) Representative asymmetry spectra (symbols) acquired in V in zero external 
magnetic field, and fits (solid lines) to the numerical dynamic Gaussian Kubo-Toyabe function.
(b) The temperature dependence of the extracted muon hop rate $\nu$ (circles).  The solid line 
is a fit to $\nu(T \geq 15$~K) using an Arrhenius function.}
\label{ZF}
\end{figure}

\section{Zero-field measurements}
Figure~\ref{ZF}(a) shows asymmetry spectra for our V sample in zero external magnetic field.
These spectra contain a 7~\% time-independent background contribution from muons
stopping outside the sample. The signal coming from muons stopping inside the sample is well
described by a numerical dynamic Gaussian Kubo-Toyabe function.\cite{Schenck:85} 
This function is characterized by a relaxation rate $\Delta$ corresponding to the
width of the internal magnetic field distribution experienced by the muons,
and a parameter $\nu$ corresponding to the hop rate of the muons in the sample.
As shown in Fig.~\ref{ZF}(b), the muon hop rate in our V crystal
decreases with decreasing temperature to $\nu \! \approx \! 0.2$~$\mu s^{-1}$ at $T = 15$~K 
($1/T \approx 0.07$~K$^{-1}$).
At $T \geq 15$~K, the data are well described by the classical Arrhenius law for thermally activated 
motion in the presence of potential barriers\cite{Schenck:85}
\begin{equation}
\nu=\nu_0\exp(-E_a/k_BT),
\label{arrhenius}
\end{equation}
where $k_B$ is Boltzmann's constant, $E_a$ is the activation energy for thermally assisted muon hopping, 
and $\nu_0$ is a constant. 
Fitting the $T \geq 15$~K data using this expression yields $\nu_0 = 1.72$~$\mu$s$^{-1}$ and 
$E_a = 4.5$~meV.
Below $T = 15$~K there is perhaps a slow increase in the hop rate $\nu$, which we 
speculate is due to quantum mechanical tunneling as observed in other metals.\cite{Storchak}

Assuming that the muon occupies an interstitial site of tetrahedral symmetry in the V crystal 
lattice, we can calculate the muon diffusivity $D_{\mu}$ from the expression\cite{Schenck:85}
\begin{equation}
D_{\mu}=\nu \frac{a^2}{24},
\label{arrhenius}
\end{equation}
where $a = 3.02$~{\AA} is the lattice constant.
This gives $D_{\mu} \approx 9.7\times10^{-16}$~m$^2$/s at $T = 2.5$~K.
Brandt and Seeger performed a thorough theoretical study of 
the effect of muon diffusion on $\mu$SR lineshapes in the vortex state.\cite{Brandt:86}  
They found that muon 
diffusion causes significant smearing of the sharp features of $n(B)$ for values of
$D_{\mu}$ greater than $\approx 10^{-3} \gamma_{\mu}|M|d^2$, where $M$ is 
the sample magnetization and $d$ is the intervortex spacing.  
Our measured muon diffusivity is several orders of magnitude smaller than this.
At $H = 1.6$~kOe, for example, we have 
$\gamma_{\mu}|M|d^2 \approx 2\times10^{-8}$~m$^2$/s, which means that 
$D_{\mu} (\approx 9.7\times10^{-16}$~m$^2$/s)~$= 5\times10^{-8} \gamma_{\mu}|M|d^2$.  
Thus muon diffusion has a negligible effect on the $\mu$SR linehapes measured here.

\section{Measurements in the vortex state}
\subsection{Comparison of fits}
To fit the $\mu$SR signals in the vortex state, the field distribution
$n(B)$ contained in the theoretical polarization function $P_x(t)$ of 
Eq.~(\ref{eq:polarization}) was generated from one of the three theoretical models 
for $B({\bf r})$ described in Sec.~II. 
For all cases, an ideal hexagonal vortex lattice was assumed. 
In addition, $P_x(t)$ was multiplied
by a Gaussian function $\exp(-\sigma^2 t^2/2)$, which is equivalent to 
convoluting $n(B)$ with the Gaussian $\exp[-(\gamma_\mu B/ \sigma)^2/2]$.
This accounts for disorder
in the vortex lattice,\cite{Brandt:88b} and the static local-field inhomogeneity
created by the large $^{51}$V nuclear dipole moments. An additional Gaussian
depolarization function was added to Eq.~(\ref{eq:asy}) to account for 
approximately 20~\% of the signal arising from muons that stopped outside the sample.  
\begin{figure}
\centering
\includegraphics[width=6.0cm, angle=270]{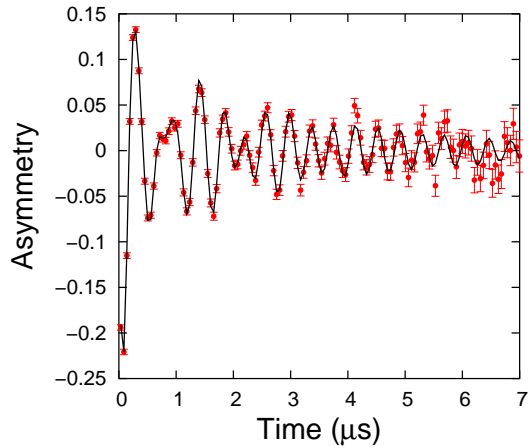}
\caption{(Color online) Time evolution of the muon spin polarization (cirlces) in V
at $H \! = \! 1.6$~kOe and $T \! = \! 2.6$~K, and a fit (solid curve) assuming
the solution of $B({\bf r})$ from the iterative GL method.}
\label{asy}
\end{figure}

\begin{figure}
\centering
\includegraphics[width=9.0cm]{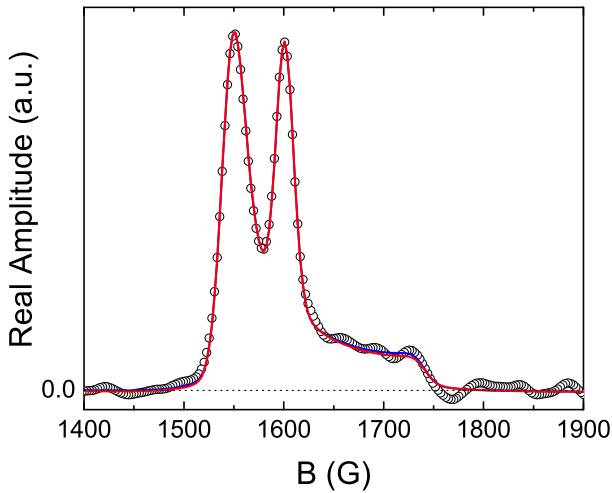}
\caption{(Color online) Fourier transforms of the muon spin precession
signal in V at $H\!=\!1.6$~kOe and $T\!=\!2.6$~K (circles), 
the fit using the iterative GL solution (blue curve) and the fit
to the analytical GL model (red curve).
The peak at 1600~G is the background signal originating from 
muons that missed the sample.}
\label{fft}
\end{figure}

\begin{figure}
\centering
\includegraphics[width=9.0cm]{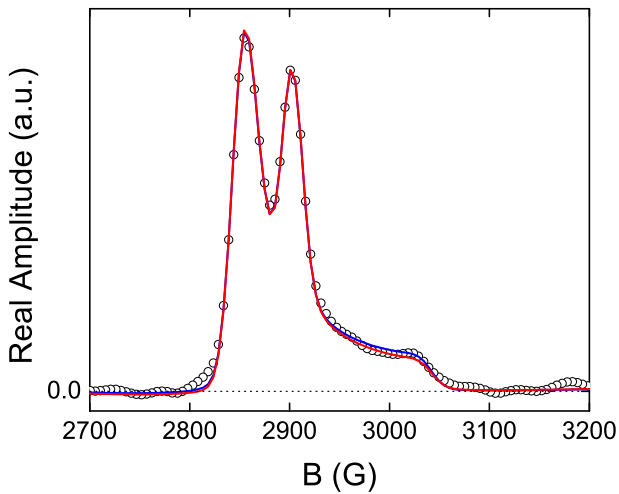}
\caption{(Color online) Fourier transforms of the muon spin precession
signal in V at $H\!=\!2.9$~kOe and $T \! = \! 0.02$~K (circles), the fit
using the iterative GL solution (blue curve) and the fit to the analytical 
GL model (red curve). 
The peak near 2900~G is the background signal originating from 
muons that missed the sample.}
\label{fft2}
\end{figure}

A typical asymmetry spectrum at $H \! = \! 1.6$~kOe and $T \! = \! 2.6$~K
is displayed in Fig.~\ref{asy}. The solid curve
through the data is a fit assuming the solution of $B({\bf r})$ obtained from
the iterative GL method.  The parameter values obtained from this fit were used to 
calculate the spatial field profiles shown in Fig.~\ref{vanad}.  
Fourier transforms of typical time domain signals
and fits to both the iterative and analytical models are shown in 
Figs.~\ref{fft} and \ref{fft2}. 
From the Fourier transforms one can see that both fits capture the main features 
of the $\mu$SR lineshape and are of high quality.
In particular, for the data in Fig.~\ref{fft} the ratio of $\chi^2$ to 
the number of degrees of freedom (NDF) is comparable, being 1.26 for the fit to the 
iterative GL solution and 1.30 for the fit to the analytical GL model.  
The values of $\lambda$, $\xi$ and $\sigma$ extracted from the two models differed by 
9~\%, 8~\% and 13~\%, respectively.  
We note that despite returning significantly different 
parameter values in some cases, fits with both models 
resulted in similar values of $\chi^2/$NDF for all of the data presented in this article.

Interestingly, the results from the two
models are in slightly better agreement at the lowest temperatures and magnetic fields. 
For example, at $H\!=\!1.2$~kOe 
and $T \! = \! 0.02$~K, the
differences in $\lambda$, $\xi$ and $\sigma$ are 7~\%, 8~\% and 2~\%, respectively.
On the other hand, at $H \! = \! 2.9$~kOe and $T \! = \! 0.02$~K, 
$\lambda$, $\xi$ and $\sigma$ differ by 10~\%, 9~\% and 3~\%, respectively.
Even so, the quality of the fits is about the same for 
both models.  This is evident from the Fourier transforms shown in Fig.~\ref{fft2}.
At high reduced field $b$, the values of $\lambda$ and $\xi$ obtained from the 
iterative GL method are likely to be more accurate, since at these high fields 
the analytical GL model is being applied outside its range of validity.

In the next section, complete results for fits 
using both the iterative and analytical GL models are presented. There we show that 
despite differences in the absolute values of $\lambda$ and $\xi$,
fits to both models yield similar temperature and magnetic field dependences for these 
length scales. In particular, 
we show that the value of $\lambda$ extrapolated to zero field is, within experimental 
uncertainty, identical for both models, and furthermore agrees well with 
Meissner state measurements on V using other experimental techniques. 

\subsection{Temperature dependences of $\lambda$ and $\xi$}

\begin{figure}
\centering
\includegraphics[width=9.0cm]{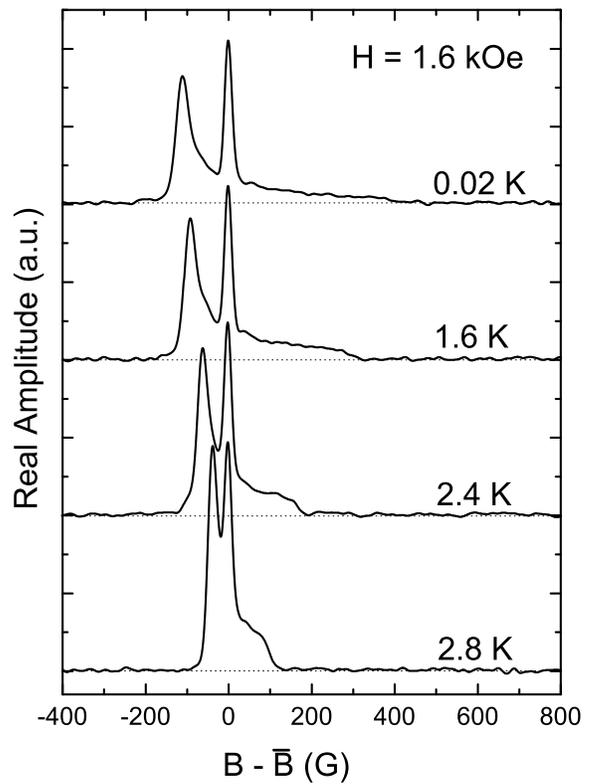}
\caption{Fourier transforms of the muon spin precession signal in V at 
$H\!=\!1.6$~kOe and $T \! < \! T_c(H)\! = \! 3.65$~K.}
\label{lineshapevsT}
\end{figure}

Fourier transforms of the muon spin precession signal from V at $H = 1.6$~kOe
and temperatures below $T_c $ are shown in Fig.~\ref{lineshapevsT}.
Magnetization measurements indicate that $T_c \! = \! 3.65$~K at 
$H = 1.6$~kOe. As the temperature is lowered, the $\mu$SR line shape broadens 
and the amplitude of the high-field ``tail'' decreases. While the high-field cutoff
is less obvious at $T \! = \! 0.02$~K, we note that the ``true'' cutoff in $n(B)$ is
smeared out by the Fourier transform.\cite{SonierRMP} 
In fact the fits in the time domain are quite sensitive to 
the high-field tail, yielding finite values for $\xi$, even at 
$T \! = \! 0.02$~K. 

\begin{figure}
\centering
\includegraphics[width=9.0cm]{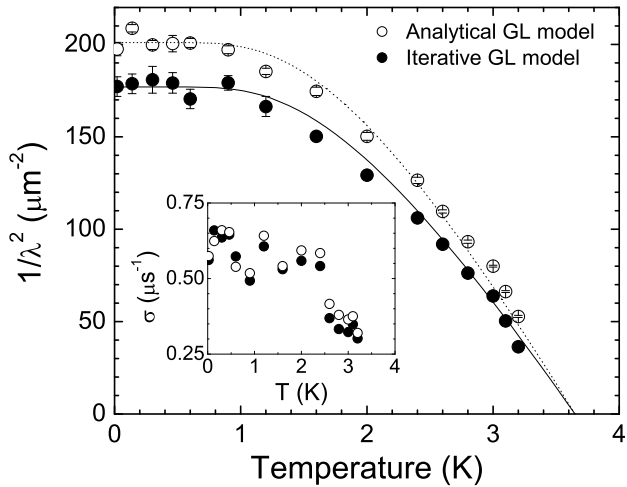} 
\caption{Temperature dependence of $1/\lambda^2$ in V
at $H\!=\!1.6$~kOe, determined from fits using the iterative and analytical GL models.
The solid and dotted curves are theoretical BCS weak-coupling predictions for 
$T_c \! = \! 3.65$~K.}
\label{lamsq}
\end{figure}

\begin{figure}
\centering
\includegraphics[width=9.0cm]{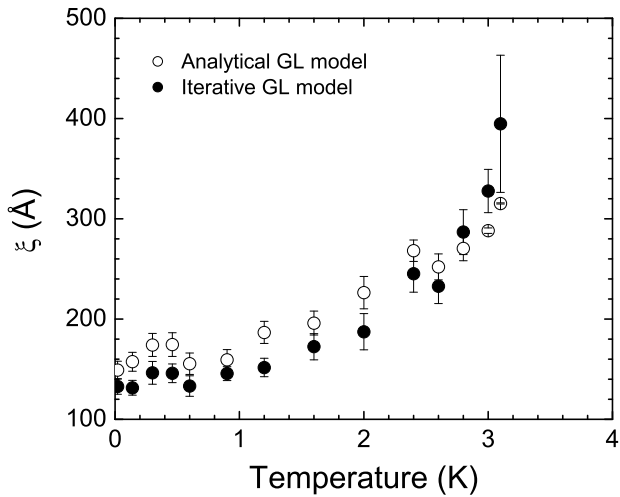}
\caption{Temperature dependence of $\xi$ in V at $H\!=\!1.6$~kOe, 
determined from fits using the iterative and anlytical GL models.}
\label{xivsT}
\end{figure}

Figures~\ref{lamsq} and \ref{xivsT} show the temperature dependences of $1/\lambda^2$, $\sigma$
and $\xi$ determined from our fits of the muon spin precession signals at $H \! = \! 1.6$~kOe,
assuming solutions for $B({\bf r})$ from both the iterative and analytical GL methods.
Despite the differences in absolute values of $1/\lambda^2$, both data sets in Fig.~\ref{lamsq} 
are well described by BCS weak-coupling $1/\lambda^2(T)$ curves\cite{Muhlschlegel:59} for
$T_c \! = \! 3.65$~K.  The inset in Fig.~\ref{lamsq} shows that both models yield similar 
values for the additional broadening parameter $\sigma$.
As $\lambda$ becomes longer 
with increasing temperature, there is a greater overlap of the vortices and a corresponding 
reduction in the pinning-induced disorder of the vortex lattice.  This is because the energy 
associated with the interaction between vortex lines depends on $\lambda$.\cite{Tinkham}
For this reason $\sigma(T)$ roughly follows $1/\lambda^2$ in Fig.~\ref{lamsq}.

The temperature dependence of the GL coherence length $\xi$ is shown in 
Fig.~\ref{xivsT}.  $\xi$ is a measure of 
the vortex core size. Recently, we have demonstrated from 
$\mu$SR and thermal conductivity
measurements on BCS superconductors\cite{Sonier:04,Callaghan:05} 
that the core size is dependent on the degree of localization of the 
quasiparticle bound core states. 
Thermal depopulation of the more spatially extended high-energy core states
results in a shrinking of the core size with decreasing temperature. This is
the so-called ``Kramer-Pesch effect'',\cite{Kramer:74} which has previously
been observed in NbSe$_2$ by $\mu$SR\cite{Sonier:97,Miller:00} and shown to
be dependent on magnetic field.\cite{Sonier:04b} In a clean 
BCS type-II superconductor the core size of an isolated vortex is expected to
be temperature independent below $T \! \sim T_c/k_F \xi_0$, where $k_F$
is the Fermi wave number and $\xi_0$ is the BCS coherence length.
We see in Fig.~\ref{xivsT} that $\xi(T)$ obtained from the fits to both models 
displays the Kramer-Pesch effect, with $\xi$ saturating below $T \approx 1$~K.
Using the free-electron expression $k_F = (3 \pi^2 n)^{1/3}$ (Ref.\cite{Kittel}) and 
$n \approx 9 \times 10^{28}$~m$^{-3}$ from Hall resistance measurements,
\cite{Hurd} we obtain $k_F \approx 1.4 \times 10^{10}$~m$^{-1}$.
Assuming the value of the superconducting coherence length  
$\xi_0 \! \approx \! 280$~{\AA} estimated from the
extrapolated zero-temperature value of $H_{c2}$,
the core size in our V crystal is therefore theoretically expected to saturate below 
$T \! \approx  \! 10^{-2}$~K.   
The premature saturation of the core size observed in Fig.~\ref{xivsT} could result 
from quasiparticle scattering by nonmagnetic impurities,\cite{Hayashi:05} but this 
is unlikely given that our sample is in the clean limit.  
It is important to note that theoretical predictions only exist for
isolated vortices.\cite{Kramer:74,Hayashi:98,Hayashi:05}  
In a lattice, the core states of nearest-neighbor vortices overlap to some
degree, and this is likely the reason why the strength of the Kramer-Pesch effect
observed by $\mu$SR weakens with increasing field.\cite{Sonier:04b}     
The delocalization of core states due to vortex-vortex interactions, and the 
corresponding reduction in the core size, also explains why the low-temperature value 
of $\xi$ in Fig.~\ref{xivsT} is much smaller than $\xi_0$.

\subsection{Magnetic field dependences of $\lambda$ and $\xi$}

\begin{figure}
\centering
\includegraphics[width=9.0cm]{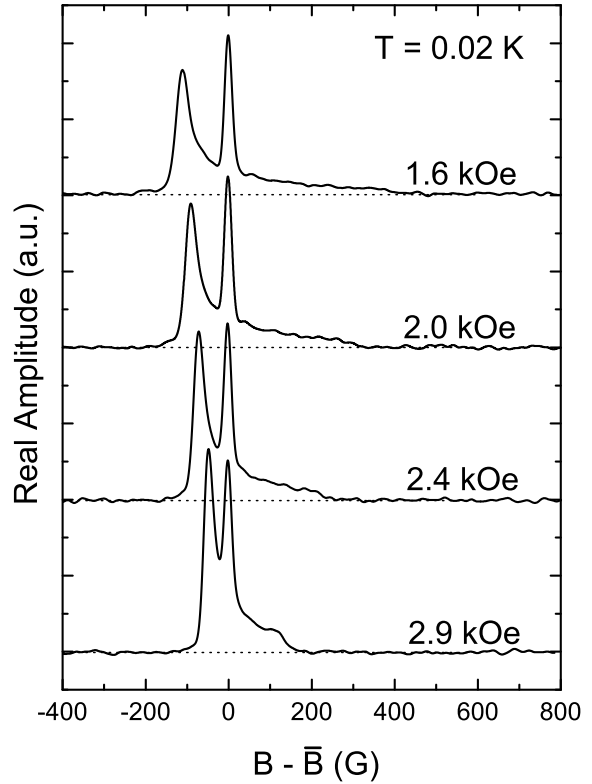}
\caption{Fourier transforms of the muon spin precession signal in V at 
$T \! = \! 0.02$~K and $H \! < \! H_{c2}$.}
\label{lineshapevsH}
\end{figure}

In Fig.~\ref{lineshapevsH}, Fourier transforms of the muon spin precession signal in V
at $T \! = \! 0.02$~K are shown for different applied magnetic fields $H \! < \! H_{c2}$.
The changes in the $\mu$SR line shape as a function of $H$ are similar to that previously
observed in NbSe$_2$ (Ref.~\cite{Sonier:97b}), and result directly from the change in vortex 
density. Increasing the vortex density reduces the internal magnetic field inhomogeneity and 
increases the degree of overlap of the wave functions of the core states
of neighboring vortices.

\begin{figure}
\centering
\includegraphics[width=9.0cm]{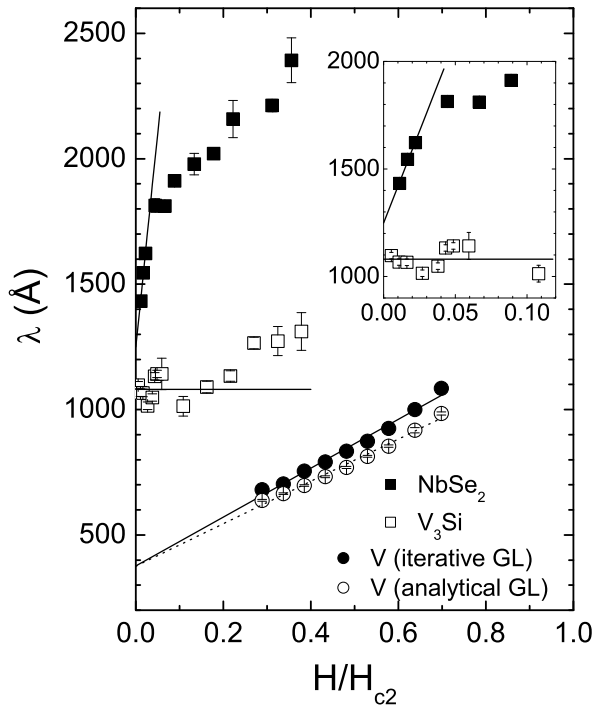}
\caption{Magnetic field dependence of $\lambda$ determined from $\mu$SR measurements
on V, V$_3$Si (Ref.~\cite{Sonier:04}) and NbSe$_2$ (Ref.~\cite{Callaghan:05}). The
straight lines are linear extrapolations of the data to zero field. 
Inset:  Blow up of the low-field data for V$_3$Si and NbSe$_2$.}
\label{lamvsB}
\end{figure}

The magnetic field dependence of $\lambda$ in V, as determined by both the analytical and 
iterative GL models, is plotted in Fig.~\ref{lamvsB}, along 
with our previously published data for V$_3$Si (Ref.~\cite{Sonier:04}) and 
NbSe$_2$ (Ref.~\cite{Callaghan:05}).  $\lambda$ from both models is seen to increase with field,
although use of the iterative model results in a slightly stronger field dependence.  
The value of $\lambda$ determined by $\mu$SR depends on the radial decay of
$B({\bf r})$ outside the vortex cores. Since the spatial field profile around a vortex core
can be significantly modifed by the delocalization of bound core states, the measured value of
$\lambda$ may be strongly dependent on field. We stress that when this is the case, 
$\lambda(H)$ is an ``effective'' length scale, which in the fits absorbs changes in 
$B({\bf r})$ due to changes in the electronic structure of the vortex lattice. This 
dominates over the weak field dependence of $\lambda$ expected for an isolated vortex 
in an $s$-wave superconductor.\cite{Bardeen:54} 
We note that the difference in slope of $\lambda$ vs $H$ between the two models in 
Fig.~\ref{lamvsB} suggests that the exact details of how these changes in electronic 
structure are absorbed by $\lambda(H)$ is model dependent.
To compare with measurements of
$\lambda$ by other experimental techniques, we have extrapolated the data for
$\lambda(H)$ to $H \! \rightarrow \! 0$~kOe. 
The zero-field extrapolated value 
of $\lambda$ in V is $376.3 \! \pm \! 22.4$~{\AA} using the iterative GL model and 
$375.9 \! \pm \! 17.0$~{\AA} using the analytical GL model.  Magnetization 
measurements\cite{Moser:82,Radebaugh:66a,Sekula:72}
have determined that $\lambda$ is in the range 374~{\AA} to 398~{\AA}, in excellent 
agreement with our results.
We see that either model for $B({\bf r})$ can be used to extract a reliable measure of 
the zero-field magnetic penetration depth.

In V$_3$Si, where the 
bound core states are highly localized at low fields,\cite{Boaknin:03} $\lambda$ is weakly 
dependent on field below $H \! \approx \! 0.2 H_{c2}$ (Ref.~\cite{Sonier:04}). The zero-field
extrapolation shown in the inset of Fig.~\ref{lamvsB} yields 
$\lambda \! = \! 1080 \!~\pm~\! 17$~{\AA}, in good agreement with the value
$\lambda \! = \! 1050$~{\AA} determined from $H_{c2}(T)$ measurements 
in Ref.~\cite{Yethiraj:05}.     
Likewise, the zero-field extrapolated value $\lambda = \! 1249 \! \pm \! 31$~{\AA} for NbSe$_2$
agrees well with the value of 1240~{\AA} obtained in Ref.~\cite{Finley:80}
from magnetization measurements. Thus $\mu$SR can be used for 
accurate measurements of the absolute value of the magnetic penetration depth in 
type-II superconductors, provided there is sufficient data to permit an accurate 
extrapolation to zero field.
This works even in the case of an unconventional superconductor.
Recently, it was shown that zero-field extrapolated values of $\lambda$ obtained 
from $\mu$SR measurements on the high temperature superconductor 
YBa$_2$Cu$_3$O$_{6+x}$ agree well with values obtained from accurate electron spin resonance 
measurements in the Meissner phase.\cite{Pereg:04} 
We note that the linear extrapolations of the data 
in Fig.~\ref{lamvsB} are continuous through the Meissner phase---which
occurs in V below $H_{c1} \! \approx \! 0.25 H_{c2}$. 

\begin{figure}
\centering
\includegraphics[width=9.0cm]{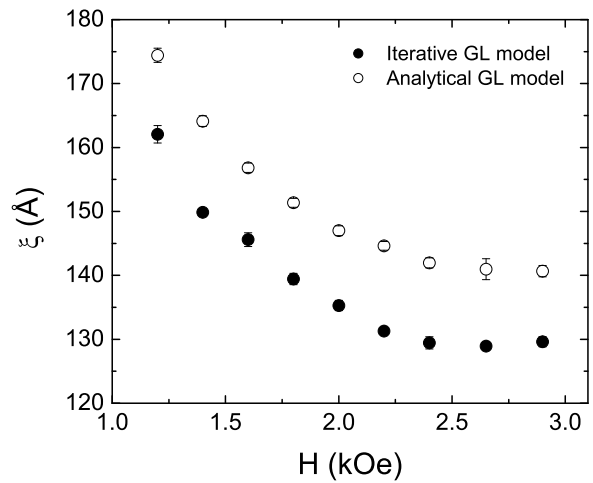}
\caption{Magnetic field dependence of $\xi$ in V at $T \! = \! 0.02$~K 
as determined from fits using the iterative and analytical GL models.}
\label{xivsH}
\end{figure}

The magnetic field dependence of $\xi$ at $T \! = \! 0.02$~K 
is plotted in Fig.~\ref{xivsH}. 
Although the analytical model yields larger values of $\xi$, both models 
display a similar field dependence over the entire field range.
Immediately above $H_{c1}$, 
the vortex core size ($\sim \! \xi$) shrinks with increasing field 
and saturates above $H \approx 2.4$~kOe.
(We note that analysis of a recent Andreev reflection spectroscopy study of niobium 
revealed a similar trend for $\xi(H)$ over the same range of reduced fields 
$0.3 \leq b \leq 0.7$.\cite{Shan:06}) 
We attribute this behavior to an increase in the overlap of the core states 
of nearest-neighbor vortices,\cite{Ichioka:99} as was found to be the case in
V$_3$Si and NbSe$_2$.\cite{Sonier:04,Callaghan:05}
Although Kogan and Zhelezina \cite{Kogan:05} have also successfully modeled the 
field dependence of the core size in V$_3$Si and NbSe$_2$ by weak-coupling BCS 
theory, their calculations assume a large GL parameter $\kappa = \lambda/\xi$ 
and hence are not applicable to V.
In an isotropic $s$-wave superconductor, the delocalization of core
states is predicted to be significant at fields above 
$B^* \! \approx \! B_{c2}/3$, although the value of this crossover 
field is reduced somewhat by anisotropy.\cite{Nakai:04}
Specific heat\cite{Radebaugh:66a} and ultrasonic attenuation\cite{Bohm:63}
measurements suggest that the anisotropy of the superconducting energy gap in V is
approximately 10~\%. According to the calculations of Ref.~\cite{Nakai:04}, significant 
delocalization of the core states and a reduction in the core size should occur 
above $B^* \approx 1.3$~kG.  Given the uncertainty in the values of $B_{c2}(T \to 0)$ 
and the anisotropy, the observed shrinking of the vortex cores at fields $H \geq 1.2$~kOe 
seems reasonable.

\begin{figure}
\centering
\includegraphics[width=9.0cm]{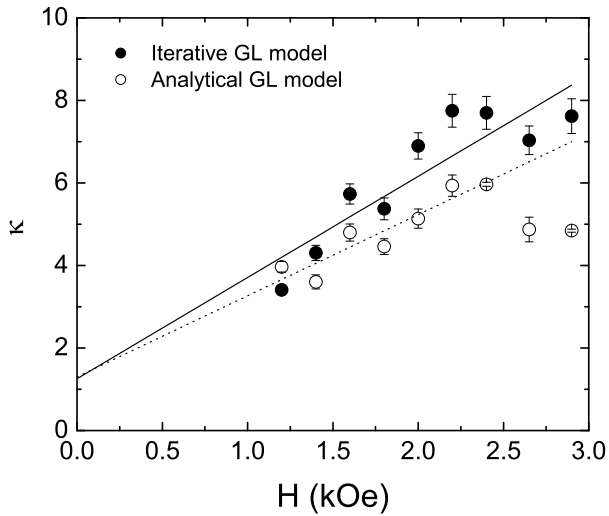}
\caption{Magnetic field dependence of $\kappa \! = \! \lambda/\xi$ 
in V at $T \! = \! 0.02$~K as determined from the iterative and 
analytical GL models.  The solid (dotted) line is a linear fit to the data 
obtained from the iterative (analytical) GL model (see main text).  
The fit to the data obtained from 
the analytical model excludes the points at $H = 2.65$~kOe and 2.9~kOe.}
\label{kappavsH}
\end{figure} 

Finally, we present the magnetic field dependence of the GL parameter $\kappa = \lambda/\xi$ 
in Fig.~\ref{kappavsH}.  
The increase in $\kappa$ with field is due to the field dependences of 
$\lambda$ and $\xi$.
The straight lines in Fig~\ref{kappavsH} show that $\kappa$ is roughly a linear function of $H$.
Significant deviations from this behavior occur for the values of $\kappa$ at the 
two highest fields determined from fits to the analytical GL model.
This is perhaps due to the fact that the analytical model breaks down at high reduced 
fields $b$.
It was found that both scatter and uncertainty in the values of $\lambda$ and $\xi$ 
were considerably reduced by fixing $\kappa$ at each field to lie on the 
straight line fits shown in Fig.~\ref{kappavsH}.  The $\xi(H)$ and $\lambda(H)$ data in 
Figs.~\ref{lamvsB} and \ref{xivsH} were obtained in this way.
The zero-field extrapolated values of $\kappa$ determined from the fits shown in 
Fig.~\ref{kappavsH} are 
$1.26 \pm 1.01$ for the iterative model and $1.30 \pm 0.64$ for the analytical model.
We note that both of these values agree well with the value $\kappa = 1.34$ 
calculated from our estimated values of $\lambda(H=0)$ and $\xi_0$. 
Also, within experimental uncertainty, both extrapolated values of $\kappa$ are comparable with that 
obtained by other experimental methods for samples of similar purity.\cite{Moser:82}

\section{Summary and Conclusions}
We have analyzed $\mu$SR measurements of the internal magnetic field
distribution in the vortex state of the low-$\kappa$ type-II 
superconductor V using both Brandt's iterative GL method\cite{Brandt:97} and 
the analytical GL model of Ref.~\cite{Yaouanc:97}.
Surprisingly, the two  models produce qualitatively similar results for both the 
temperature and field dependences of $\lambda$ and $\xi$. In particular, 
fits to each model yield low-temperature, zero-field extrapolated values of 
$\lambda$ and $\kappa$ that agree with previous measurements of these quantities 
by other techniques. We find that the largest difference between the results
using these models occurs at high fields, where the analytical GL model is being
applied outside its range of validity. The observed field dependences of   
$\lambda$ and $\xi$ in V are likely due to the delocalization of quasiparticle
core states, as has already been established in other conventional superconductors. 

We thank Prof.~E.~M.~Forgan and Prof.~E.~H.~Brandt for fruitful discussions and 
we are grateful to P.~Dosanjh and K.~Musselman for assistance 
with resistivity measurements.
The work presented here was supported by the Natural Science and 
Engineering Research Council of Canada, and the Canadian Institute for 
Advanced Research. We thank staff at TRIUMF's Centre for Molecular
and Materials Science for technical assistance with the $\mu$SR
experiments.


\begin{thebibliography}{xx}
\bibitem{SonierRMP} J.E.~Sonier, J.H.~Brewer, and R.F.~Kiefl, Rev.~Mod.~Phys. 
{\bf 72}, 769 (2000).

\bibitem{Oliveira:98} I.G.~de~Oliveira and A.M.~Thompson, Phys.~Rev~B 
{\bf 57}, 7477 (1998).

\bibitem{Clem:75} J.R.~Clem, J.~Low~Temp.~Phys. {\bf 18}, 427 (1975).

\bibitem{Yaouanc:97} A.~Yaouanc, P.~Dalmas~de~R\'{e}otier, and E.H.~Brandt, Phys.~Rev.~B 
{\bf 55}, 11107 (1997).

\bibitem{Hao:91} Z.~Hao, J.R.~Clem, M.W.~McElfresh, L.~Civale, A.P.~Malozemoff, and 
F.~Holtzberg, Phys.~Rev.~B {\bf 43}, 2844 (1991).

\bibitem{Pogosov:01} W.V.~Pogosov, K.I.~Kugel, A.L.~Rakhmanov, and E.H.~Brandt, 
Phys.~Rev.~B {\bf 64}, 064517 (2001).

\bibitem{Sonier:04} J.E.~Sonier, F.D.~Callaghan, R.I.~Miller, E.~Boaknin, L.~Taillefer, 
R.F.~Kiefl, J.H.~Brewer, K.F.~Poon, and J.D.~Brewer, Phys.~Rev.~Lett. 
{\bf 93}, 017002 (2004).

\bibitem{Callaghan:05} F.D.~Callaghan, M.~Laulajainen, C.V.~Kaiser, and J.E.~Sonier, 
Phys.~Rev.~Lett. {\bf 95}, 197001 (2005).

\bibitem{Sonier:97} J.E.~Sonier, J.H.~Brewer, R.F.~Kiefl, D.A.~Bonn, S.R.~Dunsiger, 
W.N.~Hardy, R.~Liang, W.A.~MacFarlane, R.I.~Miller, T.M.~Riseman, D.R.~Noakes, 
C.E.~Stronach, and M.F.~White,~Jr., Phys.~Rev.~Lett. {\bf 79}, 2875 (1997).

\bibitem{Sonier:99} J.E.~Sonier, J.H.~Brewer, R.F.~Kiefl, G.D.~Morris, R.I.~Miller, 
D.A.~Bonn, J.~Chakhalian, R.H.~Heffner, W.N.~Hardy, and R.~Liang, 
Phys.~Rev.~Lett. {\bf 83}, 4156 (1999).

\bibitem{Brandt:97} E.H.~Brandt, Phys.~Rev.~Lett. {\bf 78}, 2208 (1997).

\bibitem{Brandt:05} E.H.~Brandt (private communication).

\bibitem{Brandt:88} E.H.~Brandt, J.~Low~Temp.~Phys. {\bf 73}, 355 (1988).

\bibitem{Goodfellow} Goodfellow Ltd., Cambridge, UK (www.goodfellow.com).

\bibitem{Hurd} Colin~M.~Hurd, {\it The Hall Effect in Metals and Alloys} 
(Plenum Press, 1972).

\bibitem{Kittel} Charles~Kittel, {\it Introduction to Solid State Physics, Eighth Edition} 
(John Wiley and Sons, 2005).

\bibitem{Forgan:05} E.M.~Forgan (private communication).

\bibitem{Schenck:85} A.~Schenck, {\it Muon Spin Rotation Spectroscopy: Principles 
and Applications in Solid State Physics} (Adam Hilger, Bristol, England, 1985).

\bibitem{Storchak} V.G.~Storchak and N.V.~Prokof`ev, Rev.~Mod.~Phys. {\bf 70}, 929 (1998).  

\bibitem{Brandt:86} E.H.~Brandt and A.~Seeger, Adv.~Phys. {\bf 35}, 189 (1986).

\bibitem{Brandt:88b} E.H.~Brandt, Phys.~Rev.~B {\bf 37}, 2349(R) (1988).

\bibitem{Muhlschlegel:59} B.~Muhlschlegel, Z.~Phys. {\bf 155}, 313 (1959).

\bibitem{Tinkham} M.~Tinkham, {\it Introduction to Superconductivity, Second Edition} 
(McGraw-Hill, 1996).

\bibitem{Kramer:74} L.~Kramer and W.~Pesch, Z.~Phys. {\bf 269}, 59 (1974).

\bibitem{Miller:00} R.I.~Miller, R.F.~Kiefl, J.H.~Brewer, J.~Chakhalian, S.~Dunsiger, 
G.D.~Morris, J.E.~Sonier, and W.A.~MacFarlane, Phys.~Rev.~Lett. {\bf 85}, 1540 (2000).

\bibitem{Sonier:04b} J.E.~Sonier, J.~Phys.~Condens.~Matter {\bf 16}, S4499 (2004).

\bibitem{Hayashi:05} N.~Hayashi, Y.~Kato, and M.~Sigrist, 
J.~Low~Temp.~Phys. {\bf 139}, 79 (2005).

\bibitem{Hayashi:98} N.~Hayashi, T.~Isoshima, M.~Ichioka, and K.~Machida, 
Phys.~Rev.~Lett. {\bf 80}, 2921 (1998).

\bibitem{Sonier:97b} J.E.~Sonier, R.F.~Kiefl, J.H.~Brewer, J.~Chakhalian, S.R.~Dunsiger, 
W.A.~MacFarlane, R.I.~Miller, A.~Wong, G.M.~Luke, and J.W.~Brill, 
Phys.~Rev.~Lett. {\bf 79}, 1742 (1997).

\bibitem{Bardeen:54} J.~Bardeen, Phys.~Rev. {\bf 94}, 554 (1954).

\bibitem{Moser:82} E.~Moser, E.Seidl, and H.W.~Weber, J.~Low~Temp.~Phys. {\bf 49}, 585 (1982).

\bibitem{Radebaugh:66a} R.~Radebaugh and P.~H.~Keesom, Phys.~Rev. {\bf 149}, 217 (1966).

\bibitem{Sekula:72} S.T.~Sekula and R.H.~Kernohan, Phys.~Rev.~B {\bf 5}, 904 (1972).  

\bibitem{Boaknin:03} E.~Boaknin,  M.A.~Tanatar, J.~Paglione, D.~Hawthorn, F.~Ronning, 
R.W.~Hill, M.~Sutherland, L.~Taillefer, J.~Sonier, S.M.~Hayden, and J.W.~Brill, 
Phys.~Rev.~Lett. {\bf 90}, 117003 (2003).

\bibitem{Yethiraj:05} M.~Yethiraj,  D.K.~Christen, A.A.~Gapud, D.McK.~Paul, S.J.~Crowe, 
C.D.~Dewhurst, R.~Cubitt, L.~Porcar, and A.~Gurevich, 
Phys.~Rev.~B {\bf 72}, 060504 (2005).

\bibitem{Finley:80} J.J.~Finley and B.S.~Deaver~Jr., Solid~State~Commun. {\bf 36}, 493 (1980).

\bibitem{Pereg:04} T.~Pereg-Barnea, P.J.~Turner, R.~Harris, G.K.~Mullins, J.S.~Bobowski, 
M.~Raudsepp, R.~Liang, D.A.~Bonn, and W.N.~Hardy,
Phys.~Rev.~B {\bf 69}, 184513 (2004).

\bibitem{Shan:06} L.~Shan, Y.~Huang, C.~Ren, and H.H.~Wen, Phys.~Rev.~B {\bf 73}, 134508 (2006).

\bibitem{Ichioka:99} M.~Ichioka, A.~Hasegawa, and K.~Machida, Phys.~Rev.~B {\bf 59}, 184 (1999).

\bibitem{Kogan:05} V.G.~Kogan and N.V.~Zhelezina, Phys.~Rev.~B {\bf 71}, 134505 (2005).

\bibitem{Nakai:04} N.~Nakai, P.~Miranovi\'{c}, M.~Ichioka, and K.~Machida,
Phys.~Rev.~B {\bf 70}, 100503(R) (2004).

\bibitem{Bohm:63} H.V.~Bohm and N.H.~Horwitz, in {\it Proceedings of the Eighth International Conference on 
Low Temperature Physics}, edited by R.O.~Davies (Butterworth Scientific Publications Ltd., London, 1963), p.191.

\end{thebibliography}

\end{document}